\documentclass[12pt]{article}
\usepackage{t1enc}
\usepackage[latin1]{inputenc}
\usepackage{epsfig}
\usepackage{graphicx}
\usepackage[dvips]{rotating}
\usepackage{flafter}
\begin{document}

\begin{center}
\begin{Large}
The influence of an individual opinion \\
in the Sznajd Model
\end{Large}
\end{center}

\begin{center}NINA KLIETSCH\\
Institute for Theoretical Physics, Cologne University, D-50923 K\"oln, Germany
\end{center}

\begin{abstract}
The method of \textit{Damage Spreading} was used to simulate the influence that a \textit{single} persons' change of opionion has on the consensus opinion built up in a population if one assumes opinions to form according to the \textit{Sznajd Model}. The results confirm the intuitive assumption that there is hardly any chance for one person to change the consensus, that the effect of this change dies out after a certain time and its range decreases with time. The \textit{consensus times} were compared and it turned out that the consensus can be delayed or accelerated by this slight modification and that the amount of the difference in the consensus times obeys a certain power law as well as the lifetime of the effect. Moreover two \textit{scaling laws} concerning temporal and spatial aspects could be observed up to a certain size of population.
\end{abstract}

\begin{normalsize}


\end{normalsize}

\section{Introduction}
What if Hitler would have been killed as soldier during World War I ? Could this had stopped National Socialism in Germany ?\\
A certain, shocking event like the eruption of a vulcano, an airplane crash or Hitlers' death usually initializes a fraction of people to change their opinion in some way. An airplane crash for example can cause people to avoid airplanes for the next time and to travel by train. This behaviour is often a temporary precaution and is abandoned after a while. The eruption of Mount St. Helens certainly unsettles people close to the vulcano more than in Paris. This shows that the effect also decreases with increasing distance \footnote{For more empirical studies see \cite{DSsznajd2}, \cite{pohl}}. Thus, a shocking event has \textit{temporal} and \textit{spatial} impacts. While most of the studies on the effect of changed opinions on the opinion formation in a population deal with events that change \textit{several} opinions and are limited to their \textit{temporal} effects, in this paper the limiting case of a \textit{single} persons' opinion change is investigated and the often neglected spatial impact is also attached importance to.\\
In order to do this, a model of opinion formation is required. We work with the \textit{Sznajd Model} of 2000 although the results should be independent of the special choice of model \cite{fortunato}.  
\section{The Sznajd Model and the basic results}
The \textit{Sznajd Model} \cite{sznajdweron} assumes that the interacting people (\textit{agents}) are located on the places of a $L$$\times$$L$ square lattice of sidelength $L$ and have exactly one of two possible opinions. The Sznajd rule is:\\
\textbf{A pair of \textit{nearest neighbours} convinces its' six nearest neighbours of its' opinion if and only if they both share the same opinion. Otherwise the opinions of all eight involved agents remain unchanged.}\\
Here \textit{nearest neighbours} are two agents whose places share a common side.\\
In the simulation on a computer the two possible opinions are initially distributed randomly on the lattices' places. Then step by step one agent is chosen randomly as well as one of its nearest neighbours. Then the Sznajd rule is applied on this pair. This whole procedure is called a \textit{Sznajd process}. One goes through the lattice like a typewriter and at every position visited one Sznajd process is performed. One \textit{timestep} shall be over when one sweep through the lattice is made, i.e. $L^2$ Sznajd processes are performed in average.\\ 
This model is a \textit{consensus model} that always leads to a consensus. All agents share the same opinion at the end, one opinion dies out \cite{stauffer}. The consensus opinion depends critically on the initial random distribution of the opinions. If the probability $p$ for opinion $1$ is more than 50 percent, opinion $-1$ will vanish completely at the end of the simulation and vice versa. In the  case $p$=0.5 the possible consensus opinions $1$ and $-1$ are reached with the same probability 0.5.  
A FORTRAN-program simulating the Sznajd Model in the mentioned variant on a square lattice is listed in \cite{stauffer}.

\section{Damage Spreading in the Sznajd Model and the limiting case}
\begin{figure}[h!]
\centerline{\epsfig{file=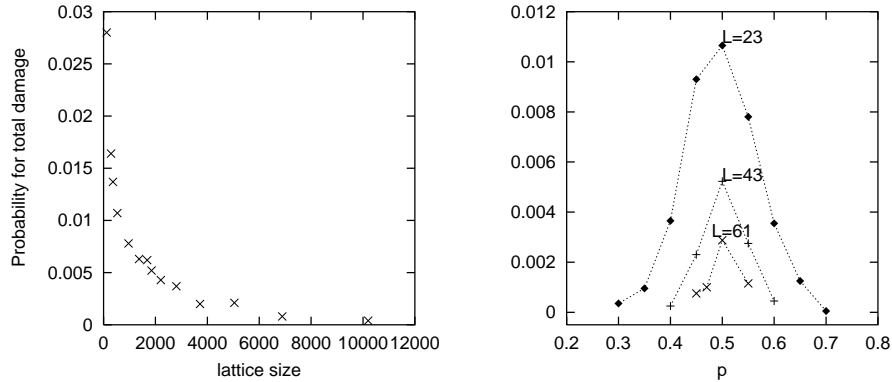, scale=0.485, angle=-90}}
\caption{Probability for total damage as a function of $p$ for different lattice sizes (right) and probability for total damage versus lattice size (left).} 
\label{tdl}
\end{figure}

The method of \textit{Damage Spreading} was first used by S. Kaufmann in biology \cite{Kaufmann} and is an useful tool to investigate the development of two systems that obey the same kind of dynamic rules and differ only in a slight modification. The strategy is very simple:\\
A replication $L_2$ of the initial system $L_1$ is created and a certain amount of elements is changed in $L_2$ (\textit{initial damage}). Then both systems develop under exactly the same conditions (i.e. the same sequence of random numbers) towards consensus and one observes the impact the initial damage has.\\
 Here the systems are two lattices, the initial damage consists in a single persons' opinion change and the rule for the dynamics is the Sznajd rule. The effect of the opinion change can be measured by the fraction of different opinions on both lattices at a certain moment and is called the \textit{damage} $D$ (consistent to the already defined initial damage) or \textit{Hamming-distance}. The damage is determined by a site-by-site comparison of both lattices while going through the lattice like a typewriter. Each site with a difference in opinions is summed up. The first used of \textit{damage spreading} in the Sznajd Model was in \cite{Bern} where more than one opinion were changed.\\
 In the Sznajd Model two different szenarios are possible:\\
The damage dies out ($D$$\rightarrow$0) or the damage spreads over the whole lattice ($D$$\rightarrow$$L^2$). The last case ends up in a reversal of the consensus opinion of the unchanged lattice and is called \textit{total damage}. When one of these two \textit{equilibrum states} is reached, the simulation can stop.\\
As already mentioned, the state of consensus is determined by the probability $p$ for opinion 1 if $p$$\neq$$0.5$. A single change of opinion does not alter this probability. Therefore the most interesting case is $p$$=$$0.5$ where total damages are most probable (see fig. \ref{tdl}) although the values are very low.
The further investigation was therefore limited to this case and it was decided to change the opinion in the center of the lattice to know the initial location of the damage. Furthermore it should be mentioned that lattice-sizes of a prime number are used to enable long periods and minimal correlations of the random-number-sequences generated by multiplication with 16807. The \textit{distance} of a damaged site to the initial damage (in the center) is measured in the \textit{Manhattan-metrics} and the \textit{range} of the damage in every timestep is the maximal distance occurring in that timestep. All values were averaged over 10,000 simulations.
\section{The influence of a single opinion} 
\begin{figure}[h!]
\includegraphics[scale=0.3,angle=-90]{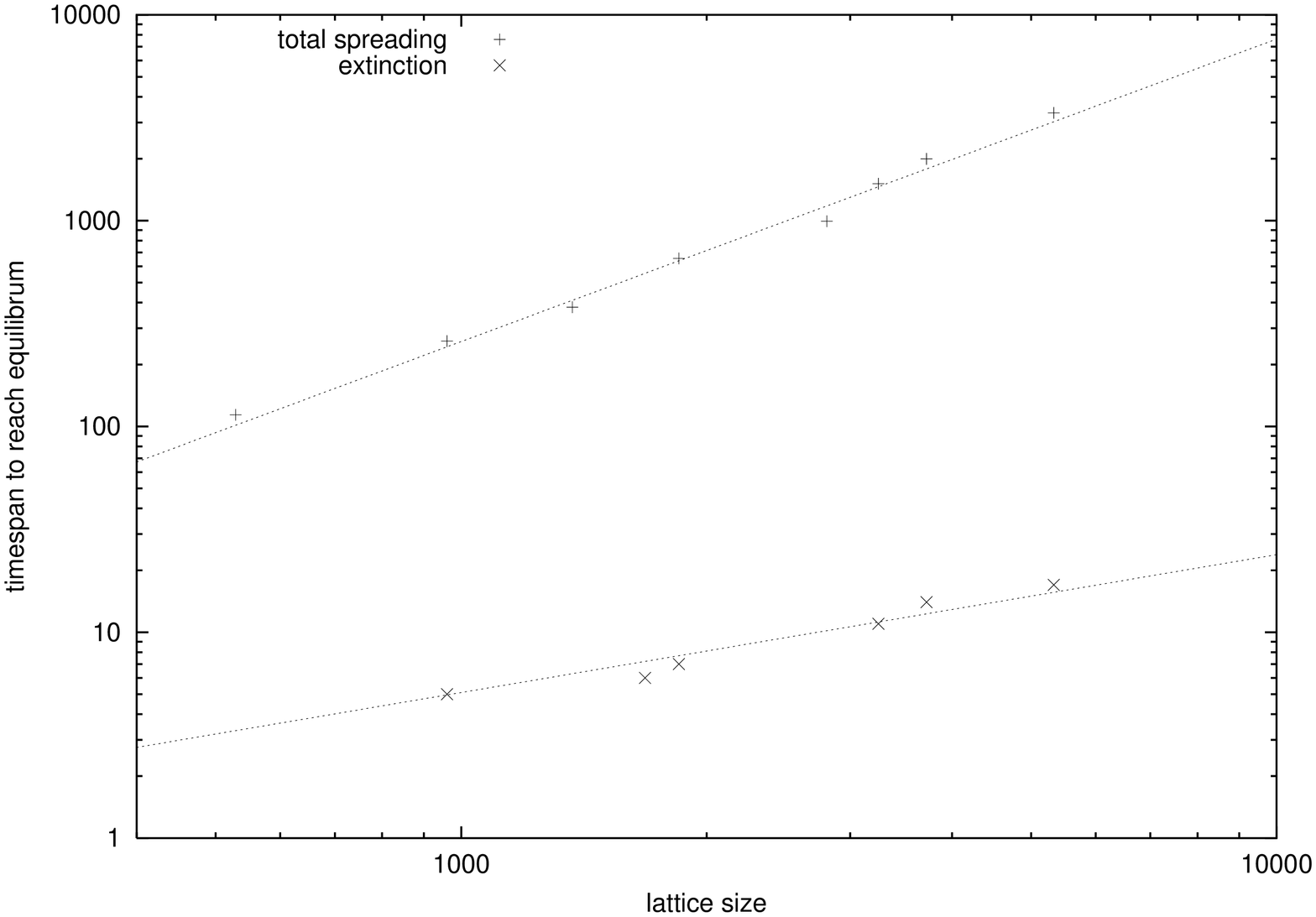}
\includegraphics[scale=0.3,angle=-90]{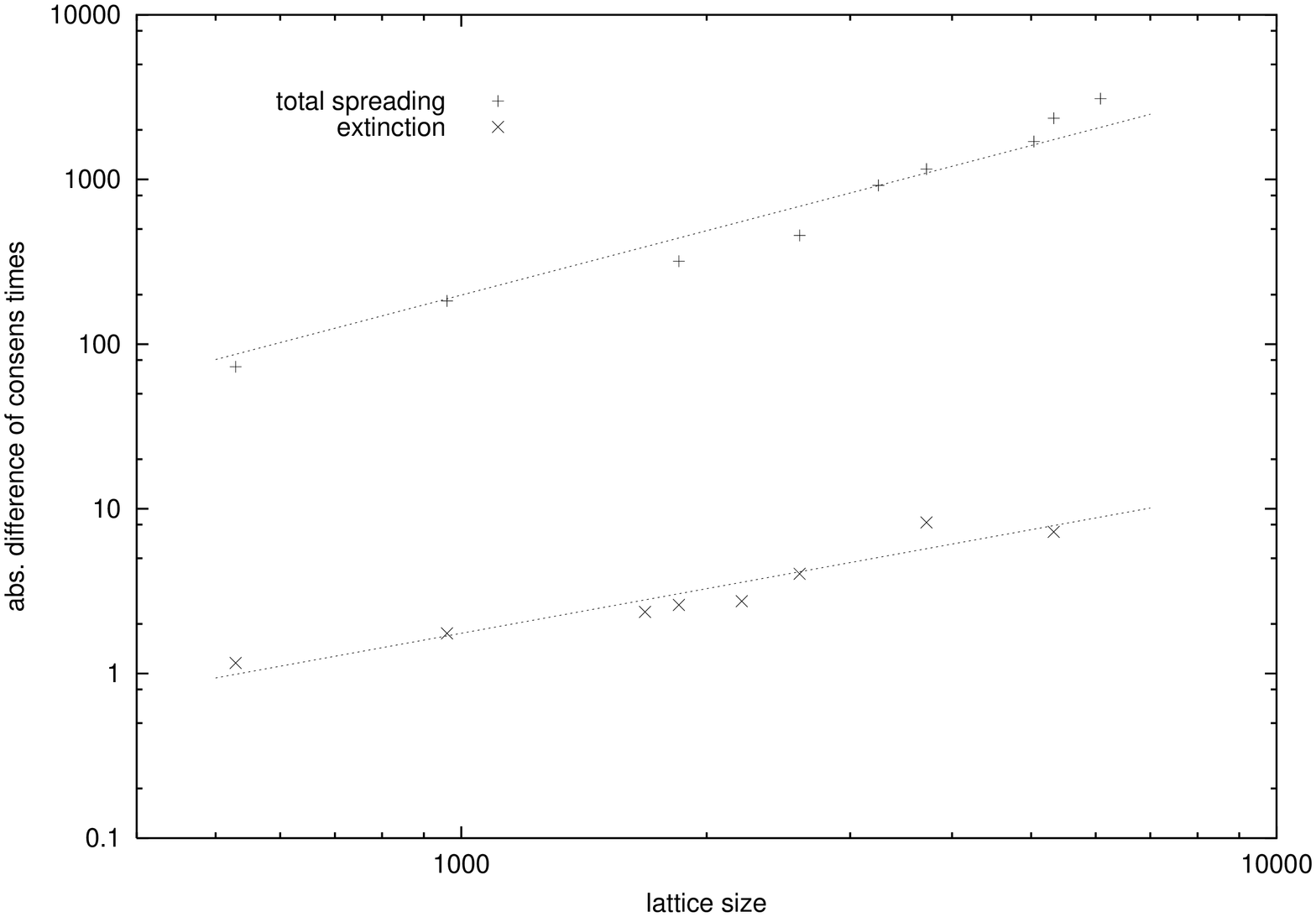}
\caption{\textit{Left}: The times needed to reach equilibrium in dependence of the lattice size $L^2$. In the log-log-plot straight lines with slopes 1.47 and 0.67 are plotted showing that the spans increase $\sim$$L^{2\cdot1.47}$ and  $\sim$$L^{2\cdot0.67}$. \textit{Right}: Absolute differences in consensus times as a function of the lattice size. $\Delta_{cons}$$\sim$ $L^{2\cdot1.3}$ for cases of total damage and $\Delta_{cons}$$\sim$$ L^{2\cdot1.1}$ for cases of extinction.} 
\label{power1}
\end{figure}

\begin{figure}[h!]
\centerline{\epsfig{file=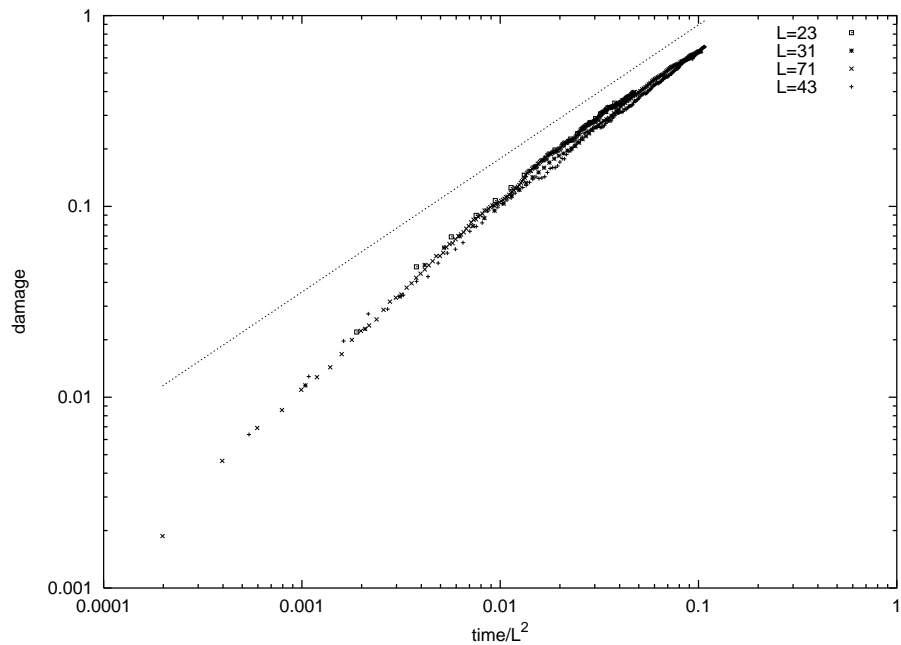, scale=0.485, angle=-90}}
\caption{The damage as a function of the scaled time $(t/L^2)$. The slope of the plotted straight line is 0.7.}  
\label{scalet}
\end{figure}

\begin{figure}[h!]
\centerline{\epsfig{file=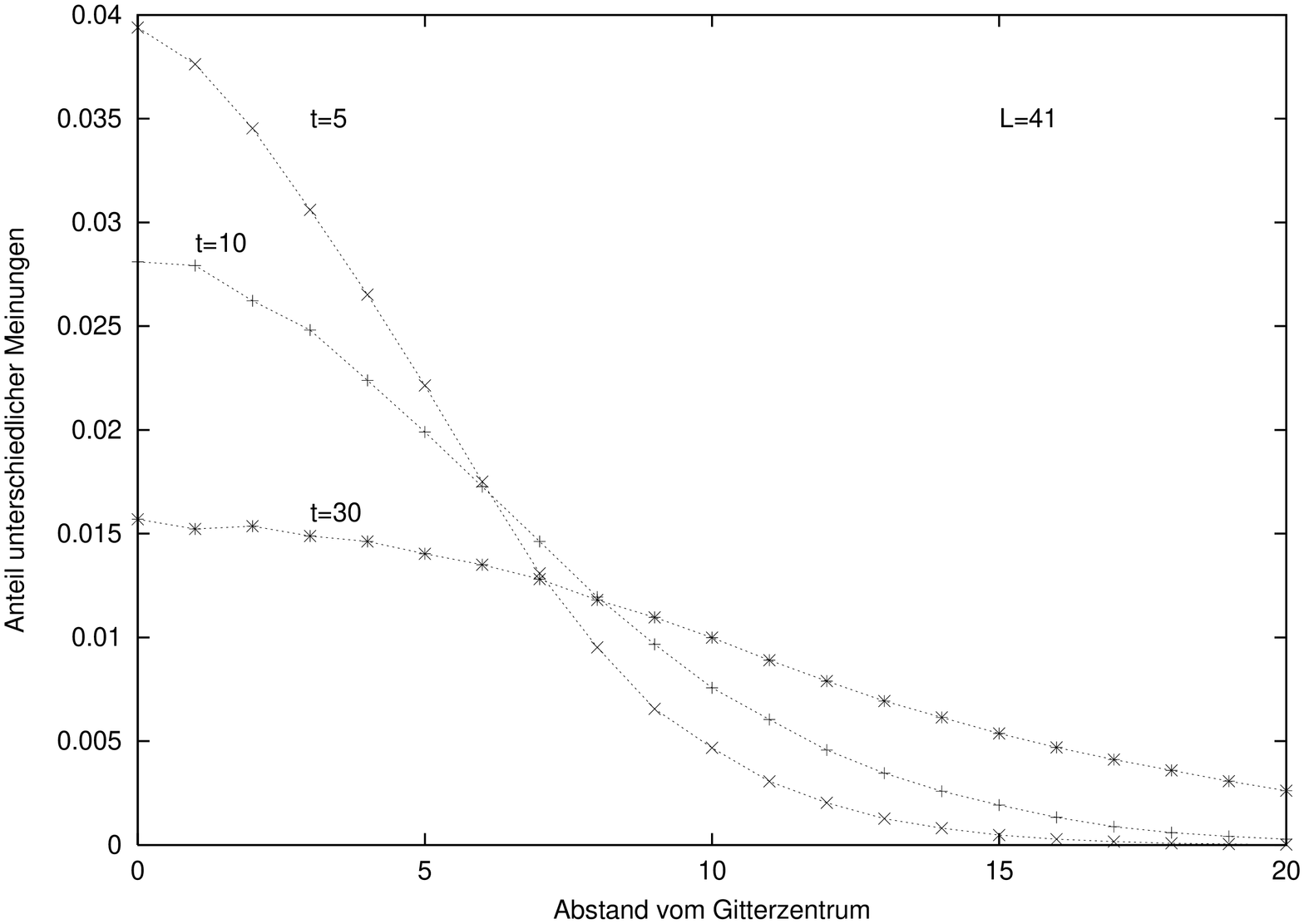, scale=0.485, angle=-90}}
\caption{Damage probability as a function of the distance at different times (on a $41$$\times$$41$-lattice).}  
\label{dmgdist}
\end{figure}

\begin{figure}[h!]
\centerline{\epsfig{file=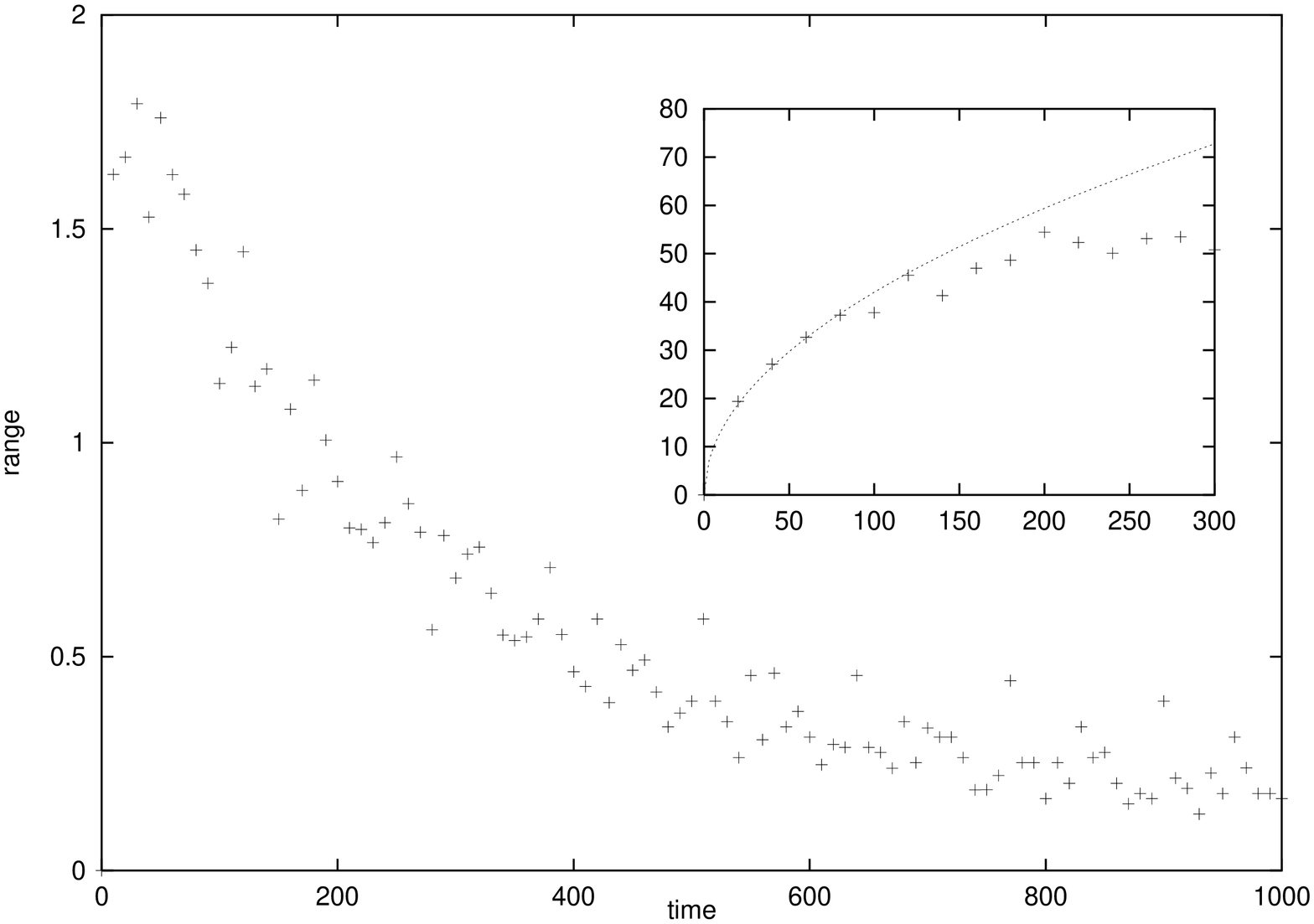, scale=0.485, angle=-90}}
\caption{The range of the damage as a function of time on a $61$$\times$$61$-lattice averaged over all samples. The inset shows that the initial spreading is proportional to $\sqrt{t}$ if the damage is averaged only over cases in which it is still alive at the considered moment. Therefore the spreading resembles a diffusion process at the very beginning.}
\label{reichweite}
\end{figure}

One result of the simulations concerning the temporal spreading of the damage is that the timespans needed to reach equilibrium (in the sense mentioned above) of the lattice size obey a power law. This could be found in cases of extinction as well as in cases of total damage (see figure \ref{power1}). Similarly the differences in times needed to reach consensus as a function of the lattice size follow a power law (figure \ref{power1}). 
This means that a consensus can be delayed or accelerated dependending on the size of the system and is a quite interesting effect: It could be sometimes quite crucial whether a consensus happens in this moment or ten years later.\\
Furthermore it is supposed that in cases of extinction the time $t$ scales with the systems' size $L^2$ and the damage is a function of this scaled time ($t/L^2$) (see figure \ref{scalet}).
 \\
The spatial investigation shows that the probability of finding a damaged site at distance $d$ from the changed opinion in the center of the lattice decreases with $d$ (figure \ref{dmgdist}).
 The dependence of the damage from the distance in the considered sizes of the system suggests a scaling law (fig. \ref{scaled}):
\begin{figure}[h!]
\centerline{\epsfig{file=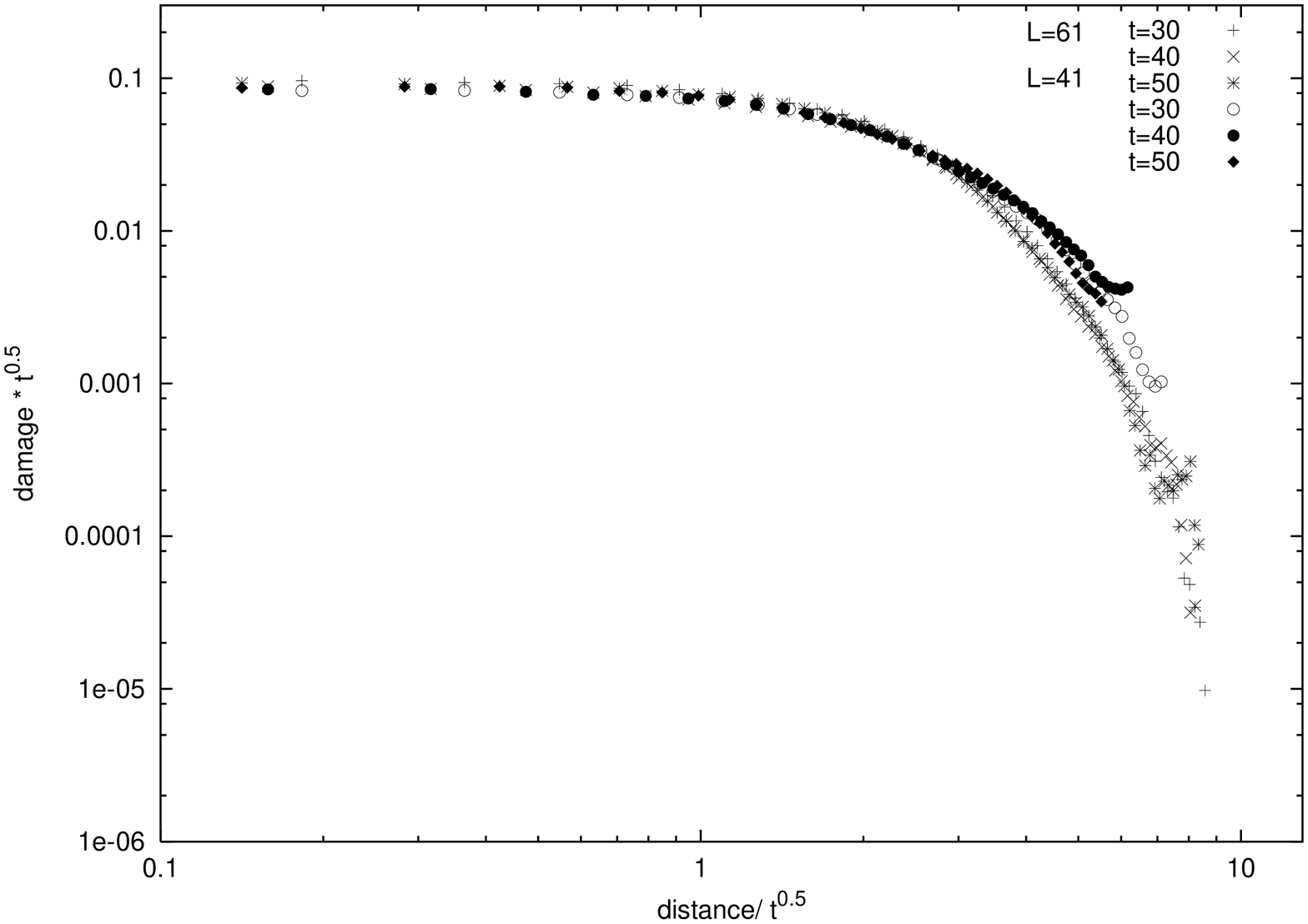, scale=0.485, angle=-90}}
\caption{Damage probability multiplied with $\sqrt{t}$ in dependence of the scaled distance $(distance/\sqrt{t})$. For $L$=101 and $t$>500 deviations appear (not shown).}  
\label{scaled}
\end{figure} 
The distance scales with $\sqrt{t}$ and the value (damage $\cdot$$\sqrt{t}$) is a function of this scaled distance ($d/\sqrt{t}$). This relation causes a comparison to a diffusion process for which such a $\sqrt{t}$-dependence is characteristic. If the range of the damage (see fig. \ref{reichweite}) is averaged over only those runs in which the damage is still alive at the moment considered, the damage increases $\sim$$\sqrt{t}$ and therefore spreads like in a diffusion process (see inset of fig. \ref{reichweite}) but finally dies out.
Finally:\\
Can the change of an opinion into the consensus-opinion of the benchmark-simulation cause a total damage ?\\
The simulations show that it is indeed possible that an opinion changed into the consensus opinion causes the opposite consensus.\\
\subsection{Conclusion}
 The results confirm the intuitive assumption that there is hardly any chance for one person to change the consensus, that the effect of this change dies out after a certain time and its range decreases with time. The \textit{consensus times} were compared and it turned out that the consensus can be delayed or accelerated by this slight modification and that the amount of the difference in the consensus times obeys a certain power law as well as the lifetime of the effect.\\
Moreover two \textit{scaling laws} concerning temporal and spatial aspects could be observed up to a certain size of population.


\begin{thebibliography}{99}
\bibitem{DSsznajd2} B. M. Roehner, D. Sornette und J.V. Andersen, \textit{Int. J. Mod. Phys. C 15,809 (2004)}
\bibitem{pohl}  R. Geipel, R. H\"arta, J. Pohl, Risiken im Mittelrheinischen Becken, \textit{Deutsche IDNDR-Reihe Nr.4}
\bibitem{fortunato} S. Fortunato, D. Stauffer, Computer Simulations of Opinions and Their Reactions to Extreme Events, in Extreme Events in Nature and Society to be edited by S. Albeverio, K. Jentsch and H.Kantz, \textit{Springer Verlag Berlin-Heidelberg}
\bibitem{sznajdweron} K. Sznajd-Weron, J. Sznajd,  Opinion evolution in a closed community, \textit{Int. J. of Mod. Phys. C 11, 1157 (2000) }
\bibitem {stauffer} D. Stauffer,  How to convince others ? Monte Carlo simulations of the Sznajd model, AIP Conf. Prof. 690, 147 (2003)
\bibitem{Kaufmann} S.A. Kaufmann, \textit{J. Theor. Biol. 22, 437 (1969)}.
\bibitem{Bern} A.T. Bernardes, U. M. S. Costa, A. D. Araujo und D. Stauffer, \textit{Int. J. Mod. Phys. C 12,159 (2001)}
\end{thebibliography}
\end{document}